# Superconducting Tunneling Spectroscopy of a Carbon Nanotube Quantum Dot

*Travis Dirks, Yung-Fu Chen, Norman O. Birge†, Nadya Mason\**

Department of Physics and Materials Research Laboratory, University of Illinois at Urbana-Champaign, Urbana, IL 61801-2902, USA

†Department of Physics and Astronomy, Michigan State University, East Lansing, MI 48824-2320, USA

ABSTRACT: We report results on superconducting tunneling spectroscopy of a carbon nanotube quantum dot. Using a three-probe technique that includes a superconducting tunnel probe, we map out changes in conductance due to band structure, excited states, and end-to-end bias. The superconducting probe allows us to observe enhanced spectroscopic features, such as robust signals of both elastic and inelastic co-tunneling. We also see evidence of inelastic scattering processes inside the quantum dot.

*Corresponding Author's Email: nadya@illinois.edu

Carbon nanotubes (CNTs) in the quantum dot regime—where confined electrons have discrete energy spectra—can demonstrate interesting physics such as electron-hole symmetry[1] and Kondo effects[2]. CNTs also form the basis of prominent schemes for implementing solid-state quantum devices,[3] such as quantum computers[4] and quantum current standards[5]. Typical studies of CNT quantum dots involve tunneling between the end contacts, a two-terminal measurement. However, there are significant advantages to performing multi-terminal measurements, which are not as dominated by a highly-variable coupling to the leads. While multi-terminal measurements on CNTs have been demonstrated with scanned probes[6] and molecular leads,[7] lithographically fabricating probes allows for the possibility of utilizing multiple probes of varying materials. For example, superconducting probes are known to enhance spectroscopic features[8,9] and enable unusual effects such as magnetic field induced tunneling of spin polarized electrons.[10] Interesting results, such as split Kondo resonances[11], and multiple Andreev reflections[8,12] have been reported in two-terminal quantum dot devices with superconducting leads. In this Letter we report three-terminal tunneling spectroscopy measurements of a CNT quantum dot, where tunneling occurs via a lithographically fabricated superconducting probe. Although contacts above or below the CNT typically create major defects in or even cut the tube,[13,14] the probe we use is largely non-invasive. We show that the superconducting tunnel probe allows otherwise invisible spectroscopic features to be observed, and also enables a more complete quantum dot spectroscopy which includes bias energy effects. The observation of such features as co-tunneling and inelastic scattering is relevant to the use of CNTs as metrological or quantum devices.

To create the device (shown in Fig. 1a), CNTs were grown via chemical vapor deposition from lithographically defined Fe catalyst islands on a degenerately doped Si wafer having 1 μm of thermally grown oxide. Scanning electron microscopy was used to locate the CNTs, which were then contacted at both ends with Pd/Au at device lengths of 1.7 μm. The entire wafer was then coated with 1.2 nm of $Al_2O_3$ via atomic layer deposition (ALD). ALD deposition of the insulator allows for manipulation of the tunnel barrier strength via layer-by-layer thickness control, and is gentle enough to not create substantial defects in the CNT. Finally, electron beam lithography was used to pattern 200-nm thick, 200-nm wide Pb tunneling probes, capped with 30 nm of In (to protect from oxidation), over the middle of the device. Devices are stable at room temperature for several weeks, but the tunnel probes do not typically survive thermal cycling. Measurements were performed in a He3 cryostat.

Conductance data show that after fabrication of the superconducting tunnel probe, the CNT remains a single, largely defect-free quantum dot. Figure 1b shows the end-to-end zero-bias conductance of a metallic device at 250 mK as a function of back gate voltage, $V_g$. The well defined Coulomb blockade peak structure occurs because of the finite energy required to add each electron to the quantum dot. The sets of four peaks are a signature of four-fold periodicity in the CNT energy levels,[15,16] due to two sub-bands (K and K') and a two-fold degenerate spin. The sub-band mismatch, δ, can be seen in the separation of groups of two within the sets of four peaks.[15] A schematic of the corresponding electronic energy level spectrum for a CNT quantum dot[1,15-17] is shown in Fig. 1c. As will be discussed, the data show that the size of the dot is consistent with the distance between the end leads. If the superconducting probe had created a significant defect in the CNT, the spacing of the Coulomb blockade peaks—particularly the four-fold periodicity[17]—would have been much more irregular.[13]

We turn now to conductance through the superconducting probe. The measurement setup is shown in Fig. 1a: a dc voltage, $V_{tunnel}$, with an ac excitation, $V_{ac}$, was applied between the superconducting tunnel probe and one end contact, and the resultant current was read out through a current preamplifier into a lock-in amplifier. A gate voltage, $V_g$, could be applied to the back of the silicon substrate while a floating bias voltage, $V_{sd}$, could be applied from end to end of the CNT. Figure 2a shows the tunneling conductance on a log scale as a function of $V_{tunnel}$ and $V_g$ at $V_{sd} = 0$.[18] The Coulomb diamond structure is similar to what has been previously observed,[1,15] with the striking exception of a zero conductance stripe that splits the diamond pattern and is consistent with the Pb superconducting gap, $2\Delta \sim 2.6$ meV. The clarity of the gap feature indicates a high-quality tunnel junction. The usual "closed" diamond pattern is evident when the superconducting probe is made normal with a magnetic field, as shown in the top inset of Fig. 2a.[19] It is also evident in Fig. 2a that the tops and bottoms of the diamonds are offset: this is because the tunneling probe also has a gating effect (we also see a weak offset between the top and bottom vertices of the end-to-end diamonds due to source-drain capacitance).

The data in Fig. 2 show four-fold periodicity similar to the end-to-end zero-bias conductance in Fig. 1b. While resonant tunneling lines make up the diamonds, excited states are also visible (denoted by red stars in Fig. 2b); these are due to conduction through an additional energy level as the tunnel bias is increased. The data is consistent with the expected stability diagram, shown in Fig. 2b. Using the diamond structure to characterize the quantum dot,[6,15] we find charging energy $U_c \sim 2.1$ meV, total capacitance $C_\Sigma \sim 80$ aF, band-mismatch $\delta \sim 0.4$ meV, CNT-backgate capacitance $C_g \sim 5.0$ aF, CNT-tunnel probe capacitance $C_{tunnel} \sim 53$ aF, CNT source plus drain capacitance $C_{sd} \sim 22$ aF, and level spacing $\Delta E \sim 1.6$ meV. The level spacing is close to that estimated by quantized energy spacing $\Delta E \sim hv_F/2L \sim 1.2$ meV for a 1.7 µm long CNT.[15]

Tunneling via a superconducting probe allows us to observe large enhancements in conductance near the superconducting gap edge. This occurs because the normalized superconducting DOS, $n_s(E) \sim \text{Re}[|E|/(E^2-\Delta^2)^{1/2}]$, is a sharply peaked function which effectively magnifies the tunneling current. In particular, we are able to observe both elastic and inelastic co-tunneling processes (blue and orange arrows in Fig. 2a), which in this case are invisible when using a normal metal probe. Co-tunneling events are higher order tunneling processes that involve the simultaneous tunneling of multiple electrons. Elastic co-tunneling, which leaves the dot in the same state, dominates at low bias and results in a conductance peak when the Fermi levels of the two contacts are aligned. With a superconducting lead, this happens when the Fermi level of the normal lead is aligned with the superconducting gap edge, yielding enhanced peaks at $V_{tunnel}=\pm\Delta/e$ (see Fig. 2c). Inelastic co-tunneling, which leaves the quantum dot in an excited state, only occurs when the bias is greater than the energy needed to put the dot in the first available excited state. Thus we see enhanced inelastic co-tunneling conductance peaks when $V_{tunnel} = \pm(\Delta+\delta)/e$ (see Fig. 2c). The transition to the inelastic regime can be sharper than the characteristic lifetime broadening of the QD states,[20] and can thus be used to get a more accurate measurement of δ than would be possible from the resonant tunneling lines. The amplitudes of co-tunneling processes also have important implications for the error rates of devices such as single electron transistors,[20,21] and set a limit on the accuracy of metrological devices such as single electron frequency-locked turnstiles, which have been proposed as current standards.[5,22] While weak inelastic co-tunneling has been previously observed in CNTs,[15,16] weak elastic co-tunneling in CNTs has only recently been seen in a two terminal device with superconducting leads.[8] The robust signals allow us to measure the elastic and inelastic co-tunneling currents at $V_{sd} = 0$ as $I_{el\text{-}co} \sim 3.7$ pA and $I_{in\text{-}co} \sim 11$ pA, respectively. The corresponding electron co-tunneling rates are $\Gamma_{el\text{-}co} \sim 2.3 \times 10^7$ s$^{-1}$ and $\Gamma_{in\text{-}co} \sim 7 \times 10^7$ s$^{-1}$. While the magnitudes of the tunneling rates depend on the DOS of the leads, the ratio of elastic to inelastic tunneling should be independent of the leads.[21] It is important to note that when the tunnel probe is made normal by a magnetic field (inset of Fig. 2a), we do not see any co-tunneling features in the Coulomb diamonds.

In addition to the observation of enhanced spectroscopic features, the three-terminal measurement allows us to directly determine the effect of end-to-end bias on the quantum dot spectrum. Figure 3a shows tunneling differential conductance from the superconducting tunneling probe to the CNT on a log scale as a function of $V_{tunnel}$ and $V_g$ while $V_{sd} = 0.8$ mV is applied across the ends. The features are similar to those for $V_{sd} = 0$ (Fig. 2a), which indicates that the energy spectrum of the dot is largely unchanged. However, another set of peaks, separated by the Pb gap energy but offset by $V_{sd}$, also appears (see red lines in Fig. 3b). These additional conduction lines show up when energy states of the CNT align with the Fermi level of the left end contact at $E = -eV_{sd}$,[23] demonstrating that the end-to-end bias can be spectroscopically determined. The resonant tunneling to both end leads through the same energy level is separated by $V_{sd}$ in the vertical direction and $(C_\Sigma/C_g)V_{sd}$ in the horizontal direction, where $C_\Sigma$ is the total capacitance of the nanotube. From this we find $C_g \sim 6.4$ aF, which agrees well with the value from the slopes of resonant tunneling lines.

When a bias is applied across the ends of the CNT we observe conductance inside the superconducting gap (see Fig. 3c), even though $V_{sd} = 0.8$ mV is smaller than the gap energy of $\sim 2\Delta/e = 2.6$ mV. The conductance in the gap is surprising since it should be suppressed exponentially,[24] and is not observed when $V_{sd} = 0$ (see Fig. 2a). It is possible that a finite source-drain bias across the tube enhances the inelastic scattering of electrons, creating excited electrons and holes that can tunnel above and below the gap, respectively, and thus create a non-zero tunnel current, $I_{sc}$. From Fig. 3c we find $I_{sc} \sim 4 - 6$ pA, which sets a lower bound on the inelastic scattering rate $\Gamma_{in}$ of $\Gamma_{in} > I_{sc}/e \sim 2.5 \times 10^7 - 3.78 \times 10^7$ s$^{-1}$. This scattering rate is typically estimated experimentally via level broadening, which often only gives an upper bound because of thermal broadening effects. The mechanism for the enhanced scattering remains unknown and will be investigated in the future.

In summary, we have described the fabrication and measurement of a device consisting of a noninvasive superconducting tunnel probe over the middle of a clean, contacted CNT quantum dot. We showed how the use of a superconducting probe enhanced tunneling signals, and how spectroscopy using this three-terminal device allowed the effects of bias to be determined. We were able to observe both elastic and inelastic co-tunneling, as well as signatures of inelastic scattering. This more complete characterization of a CNT quantum dot opens the door to a better understanding of the mechanisms behind weak, second-order processes in such systems, and allows for a better assessment of their use in practical devices such as single-electron transistors.

Acknowledgements: This research was funded by the NSF under grant DMR-0644674 (TD, YFC, NM) and DMR-0705213 (NOB), and partly carried out in the UIUC Center for Microanalysis of Materials (partially supported by the DOE under DE-FG02-07ER46453 and DE-FG02-07ER46471). We thank Nayana Shah for insightful discussions.

**Figure Caption List:**

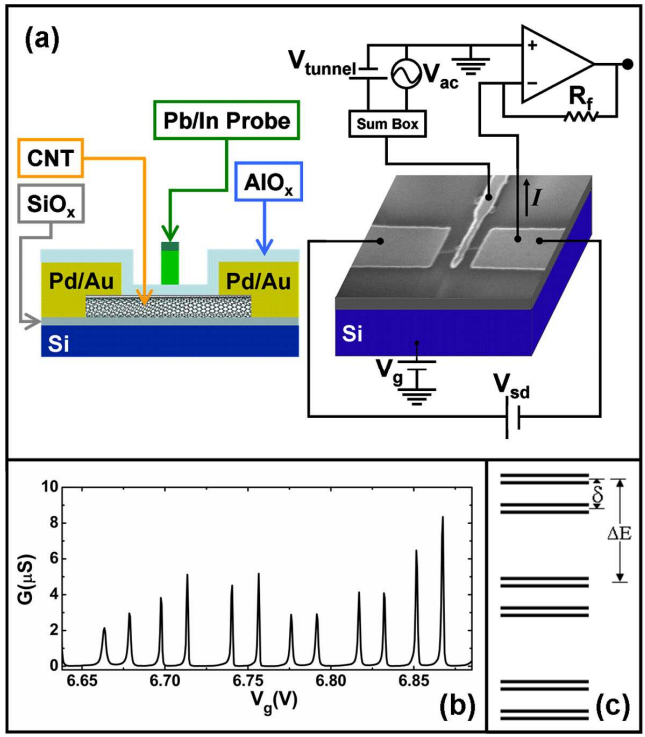

Figure 1. (a) On left, SEM image of a typical device, with diagram of the measurement circuit. On right, side view of device geometry. (b) End-to-end conductance as a function of back gate voltage. (c) Expected energy level spectrum of CNT quantum dot, showing sub-band mismatch $\delta$ and energy level spacing $\Delta E$.

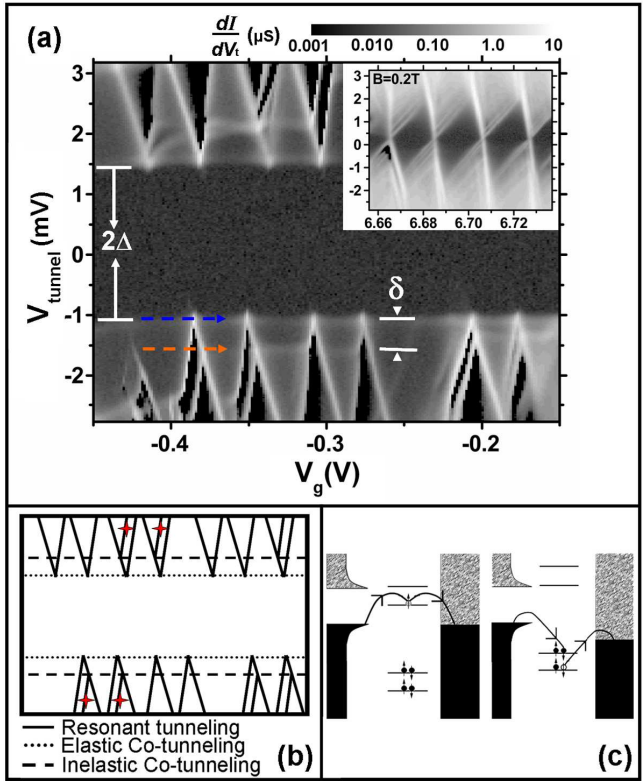

Figure 2. (a) Differential conductance between the superconducting tunnel probe and an end lead as a function of tunnel bias and back gate voltage (with end-to-end bias $V_{sd}=0$). The Pb superconducting gap, $\Delta$, and the band mismatch, $\delta$, are labeled. Blue and orange arrows point to signals of elastic and inelastic co-tunneling, respectively. Inset: Similar measurement but with an applied magnetic field, showing that the co-tunneling lines are absent when superconductivity is destroyed. (b) Expected stability diagram. Red stars indicate excited states. (c) Schematic of one of the possible elastic (left) and inelastic (right) co-tunneling processes.

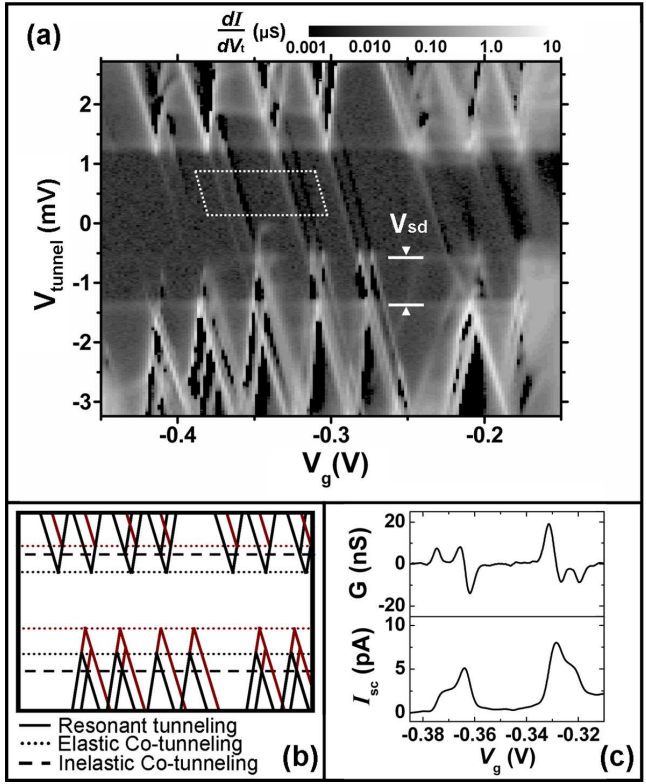

Figure 3. (a) Differential conductance between the superconducting tunnel probe and an end lead as a function of tunnel bias and back gate with $V_{sd}$ = 0.8mV applied between the end leads. Dotted box indicates data used in part C. Smeared diamonds on the right are due to a lowering of the lead tunnel barriers with gate voltage (an open dot regime). (b) Expected stability diagram. Red lines show new features expected at finite source-drain voltage. (c) Horizontal cut through some of the features inside the gap in A, with data averaged over bias range within dotted box to minimize noise, showing conductance (top) and derived current (bottom) inside the gap (cut shown on linear plot of A, since negative signals were shown as zero in log plot).

consistently observed co-tunneling that disappeared when the probe was made normal with a field.